\documentclass[aps,prl,reprint,groupedaddress]{revtex4-2}
\usepackage{graphicx}
\usepackage{epstopdf, epsfig}

\usepackage{XiGrpCommon}

\graphicspath{{figs/}}
\usepackage{CJKutf8}

\begin{document}

% \title{Linear and nonlinear pathways to turbulence in viscoelastic plane Poiseuille flow}
\title{Direct transition to elastoinertial turbulence from a linear instability in channel flow}

\author{Lu Zhu}
\author{Li Xi}
\email[corresponding author, Email: ]{xili@mcmaster.ca}
\homepage[]{http://www.xiresearch.org}
\affiliation{Department of Chemical Engineering, McMaster University, Hamilton, Ontario L8S 4L7, Canada}
\date{\today}

\begin{abstract}
For decades, transition to turbulence in viscoelastic parallel shear flows was believed to require nonlinear instabilities. We provide numerical evidences for a new wall-mode linear instability that directly triggers the transition to elastoinertial turbulence in channel flow. The instability is 2D but 3D features become important as nonlinear effects grow. With larger disturbances, direct transition to nonlinear instabilities, leading to turbulent states of both inertial and elastoinertial natures, are observed.
% For decades, viscoelastic flows induced by polymers are considered to be linear stable in rectilinear shear flows. Althrough this opinion is challenged by recent experimental and theoretical evidence, it is far from been concluded. In this study, we provide new evidence for the existence of linear instability in viscoelastic flows from direct numerical simulation (DNS). A linear transition pathway is uncovered which experiences an exponential growth of infinitesimal perturbations, a nonlinear bursting to elastoinertial turbulence (EIT), and a second bursting to inertia-driven turbulence (IDT). We found that this linear instability has 2D nature and is rely on elasticity and inertia. It persists along with the nonlinear inertia-driven and elastoinertial instabilities in different parameter regimes that form an integrated map for the laminar-turbulence (L-T) transition of the viscoelastic flow.

\end{abstract}

\maketitle

% \begin{document} % for word count only

% \section{Introduction}
In Newtonian fluids, turbulence kicks in when fluid inertia overpowers viscous dissipation to trigger flow instabilities.
% upon a critical Reynolds number $\Rey$.
The transition is nonlinear in pipe flow, for which a finite-amplitude disturbance is required.
In channel flow, although the laminar state becomes linearly unstable at the Reynolds number $\mathrm{Re}=5772$, where bifurcation to the Tollmien-Schlichting (TS) wave occurs~\citep{Orszag_JFM1971}, transition to turbulence occurs at a much lower $\mathrm{Re}_\text{c}\approx 1000$~\citep{Nishioka_Asai_JFM1985}, highlighting the essential role of nonlinear instabilities.

Dilute polymer solutions are viscoelastic. Their turbulence can show much lower friction drag, which is known as the drag reduction (DR) phenomenon~\citep{Virk_AIChEJ1975,Xi_POF2019}.
In those flows, according to the most common explanation, turbulence is still driven by inertia (inertia-driven turbulence -- IDT) but its fluctuations are suppressed by polymer stress~\citep{Li_Graham_JFM2006,Kim_Adrian_PRL2008,Zhu_Xi_JNNFM2018,Zhu_Xi_POF2019}.
% This mechanism implies that the laminar-turbulent (LT) transition would remain nonlinear and only become delayed (to higher $\mathrm{Re}_\text{c}$) compared with Newtonian flow. Delayed transition was indeed widely observed in experiments~\citep{Giles_Pettit_Nature1967,Draad_Nieuwstadt_JFM1998,Escudier_Presti_JNNFM1999}.
A different mechanism emerged recently with the discovery of the so-called elastoinertial turbulence (EIT) where both inertia and elasticity support turbulent instabilities~\citep{Samanta_Hof_PNAS2013}.
EIT dominates at the high-elasticity limit and is critical for answering the most important questions in the field, including the maximum drag reduction (MDR) phenomenon~\citep{Choueiri_Hof_PRL2018,Zhu_Xi_PRFluids2021}.
Transition to EIT can occur at lower $\mathrm{Re}_\text{c}$ (early transition) than the laminar-turbulent transition in Newtonian flow~\citep{Choueiri_Hof_PRL2018,Chandra_Shankar_JFM2020}.
% Recent discovery of the so-called elastoinertial turbulence (EIT) -- a new type of instability where both inertia and elasticity support turbulent instabilities~\citep{Samanta_Hof_PNAS2013} -- reveals a different DR mechanism that becomes important for highly-elastic fluids.
% Such fluids can show early transition in which turbulence (EIT in this case) occurs at lower $\mathrm{Re}_\text{c}$ than Newtonian flow~\citep{Choueiri_Hof_PRL2018,Chandra_Shankar_JFM2020}.
This reignited the question of whether turbulent transition can be triggered by a linear instability in viscoelastic parallel shear flow.

The answer so far has been no.
Although viscoelastic fluids can show turbulence-like flow instabilities even at vanishingly low $\mathrm{Re}$ (the purely-elastic limit)~\citep{Groisman_Steinberg_Nature2000}, 
the best known linear instability mechanisms require curved streamlines~\citep{Pakdel_McKinley_PRL1996,Xi_Graham_JFM2009}.
For parallel shear flows, the laminar state was long believed to be linearly stable~\citep{Morozov_vanSaarloos_PhysRep2007,Ho_Denn_JNNFM1977}.
Indeed, earlier search of purely elastic turbulence in those flows focused on nonlinear instabilities~\citep{Pan_Arratia_PRL2013,Qin_Arratia_PRL2019}.

Recent discovery of a center-mode (CM) linear instability in viscoelastic pipe and channel flows~\citep{Garg_Subramanian_PRL2018,Khalid_Shankar_JFM2021} has shaken this view. The instability is inertioelastic (finite $\mathrm{Re}$ is required) but, at least in channel flow, it continues smoothly to a purely-elastic ($\mathrm{Re}=0$) instability~\citep{Khalid_Shankar_PRL2021}.
EIT is probably not directly triggered by that instability.
% Contrary to what many may intuitively expect, EIT is unlikely triggered directly by this instability.
First, the CM instability is only found for a limited parameter region and requires extremely high Weissenberg number $\mathrm{Wi}$ for dilute solutions. Its parameter domain does not match where EIT is typically found.
Second, EIT structures grow from the wall regions, for which a wall-mode (WM) instability would be more intuitive~\citep{Zhu_Xi_JNNFM2020}.
However, the instability connects subcritically to a saddle-node bifurcation with an upper-branch solution resembling the so-called arrowhead structure in EIT~\citep{Page_Kerswell_PRL2020}.
For highly elastic fluids at low $\mathrm{Re}$, recent experiments observed similar flow patterns as the unstable eigenmode of the instability, which supports this subcritical connection~\citep{Choueiri_Hof_PNAS2021}.
In a competing theory, transition to EIT was linked with the so-called viscoelastic nonlinear TS wave~\citep{Shekar_Graham_JFM2020}, which, at least at high $\mathrm{Re}$, continues to the Newtonian TS wave~\citep{Shekar_Graham_PRFluids2021}.
In both scenarios, transition to EIT is believed to be subcritical and requires sufficiently large disturbances to trigger nonlinear effects.
Moreover, those studies only focused on 2D flows. Although a form of EIT does exist in 2D, key EIT dynamics important for MDR is only captured in 3D~\citep{Zhu_Xi_PRFluids2021}.

We use direct numerical simulation (DNS) to explore different transition scenarios in plane Poiseuille flow, including transition to both IDT and EIT. %EIT and the classical inertia-driven turbulence (IDT).
We report that EIT can be triggered directly from a WM linear instability in both 2D and 3D flows.
With varying $\mathrm{Re}$, $\mathrm{Wi}$, and disturbance magnitude, different nonlinear transition scenarios are also observed.

DNS is performed with a fixed pressure gradient along the $x$ direction.
Two no-slip parallel walls confine the flow in the $y$ direction. Periodic boundaries are applied to $x$ and $z$ directions.
The half-channel height $l$ and Newtonian laminar centerline velocity $U_{\mathrm{CL}}$ are used as the characteristic length and velocity, respectively, and $l/U_{\mathrm{CL}}$ and $\rho U_{\mathrm{CL}}^2$ are used to scale time $t$ and pressure $p$ ($\rho$ is fluid density).
DNS seeks the time-dependent solution of the Navier-Stokes equation coupled with the FENE-P constitutive equation~\citep{Bird_Curtis_1987}:
\begin{eqnarray}
\label{eq:ns:mom}%
\frac{D \mbf{v}}{D t} =
- \mbf{\nabla}p + \frac{\beta}{\mathrm{Re}} \nabla^{2}\mbf{v} 
+ \frac{2\left(1 -\beta\right)}{\mathrm{Re}\mathrm{Wi}}\left(\mbf{\nabla} \cdot
\mbf{\tau}_p\right),%
\\
\label{eq:ns:mass}%
\mbf{\nabla} \cdot \mbf{v} = 0,%
\\
\label{eq:fenep:conf}
\frac{\mathrm{Wi}}{2} \left(
\frac{D\mbf{\alpha}}{D t} 
-
\mbf{\alpha} \cdot \mbf{\nabla v} - \left( \mbf{\alpha} \cdot \mbf{\nabla v}
\right)^{\mathrm{T}} \right)
= -\frac{b}{b+5}\mbf{\tau}_p,
\\
\label{eq:fenep:stress}%
\mbf{\tau}_p = \frac{b + 5}{b} \left( \frac{\mbf{\alpha}}{1 -
	\frac{\mathrm{tr}(\mbf{\alpha})}{b}} -\left(\frac{b}{b + 2} \right) \mbf{\delta}
\right),%
\end{eqnarray}
where $\mbf\alpha$ and $\mbf\tau_\text{p}$ are the polymer conformation and stress tensors, respectively.
Nondimensional parameters include: $\mathrm{Re}\equiv\rho U_\text{CL}l/\eta$, $\mathrm{Wi}\equiv 2\lambda U_\text{CL}/l$, $\beta\equiv\eta_\text{s}/\eta$, and $b\equiv\max(\mathrm{tr}(\alpha))$ (where $\eta$ and $\eta_\text{s}$ are the fluid and solvent viscosity, respectively, and $\lambda$ denotes the polymer relaxation time). Polymer concentration is proportional to $1-\beta$, $\mathrm{tr}(\alpha)$ is proportional to the square of polymer end-to-end distance, and $b$ defines the finite extensibility of polymer chains.
We fix $\beta=0.97$ and $b=5000$ in this study.
A hybrid method combining
a total variation diminishing (TVD) finite difference scheme for the $\mbf v\cdot\mbf\nabla\mbf\alpha$ term with pseudospectral discretization of all other terms is adopted, in which artificial diffusion is not used~\citep{Zhu_Xi_JNNFM2020,Zhu_PhD2019}.
% Here, a TVD (total variation diminishing) finite difference scheme~\citep{zhang2015review} is utilized to discretize the $\mbf v\cdot\mbf\nabla\mbf\alpha$ term while pseudo-spectral scheme for the rest. 
% Time integration uses a third-order semi-implicit backward-differentiation/Adams-Bashforth scheme~\citep{Peyret_2002}.
DNS at different $\mathrm{Re}$ is performed with a fixed domain size $L_x\times L_y\times L_z=8.485\times2\times 2.711$ ($L_x\times L_y=8.485\times2$ for 2D).
For 3D, the numerical mesh is $N_x\times N_y\times N_z=256\times 131\times 142$ and time step is $\delta t=0.005$ and, for 2D, $N_x\times N_y=1280\times 369$ and $\delta t=0.001$.
% , which is comparable to recent EIT study~\citep{Samanta_Hof_PNAS2013,Dubief_Terrapon_POF2013,Sid_Terrapon_PRFluids2018,Shekar_Graham_PRL2019,zhu2021nonasymptotic}.
The numerical resolution and procedure were thoroughly validated in our earlier studies~\citep{Zhu_Xi_JNNFM2020,Zhu_Xi_PRFluids2021}.

An initial velocity disturbance
\begin{gather}
\label{eq:perturb3D_2}%
	(v_x^\dagger,v_y^\dagger,v_z^\dagger)|_{t=0} = \left(
		% -\frac{\partial \Psi}{\partial y}\sin\theta,
		% the sin is changed to cos, in comparison with the literature
		% otherwise in 2D, it only gives non-zero vz
		-\frac{\partial \Psi}{\partial y}\cos\theta,
		\frac{\partial \Psi}{\partial z^\prime},
		-\frac{\partial \Psi}{\partial y}\cos\theta
	\right)
\end{gather}
is superposed on the viscoelastic laminar base flow to trigger instabilities. (Hereinafter, $\dagger$ indicates deviation from the base flow.)
The disturbance stream function is 
\begin{gather}
\label{eq:perturb3D_1}%
	\Psi = Kf(y)\left(\frac{x^\prime}{l_x}\right)z^\prime
  		\exp\left[
			-\left(\frac{x^\prime}{l_x}\right)^2
			-\left(\frac{z^\prime}{l_z}\right)^2
		\right],
\end{gather}
where $l_x=l_z=2$ sets its geometric size, $\theta=0$ sets its orientation, and
\begin{gather}
\label{eq:perturb3D_3}%
  (x^\prime,z^\prime) = (x\cos\theta - z\sin\theta,x\sin\theta + z\cos\theta).
\end{gather}
The Cartesian coordinates $(x,y,z)$ are defined with the channel center as the origin.
The function
\begin{gather}
\label{eq:perturb3D_4}%
\begin{split}
	f(y) =& (1+y)^2\exp\left(\beta_f(1+y)^2\right)\\
 		  & -(1-y)^2\exp\left(\beta_f(1-y)^2\right)
\end{split}
\end{gather}
adjusts the wall-normal dependence of the disturbance. With $\beta_f=-10$ used in this study, $f(y)$ peaks at $y\approx \pm 0.7$.
$K$ adjusts the disturbance magnitude. At $K=1$, the maximum velocity disturbance is $\mathcal{O}(\num{e-2})$. % LZ: The maximum magnitude of vx, vy, vz for K=1 are 0.017, 0.016 and 0.017
The disturbance is similar to that commonly used in bypass transition studies~\citep{Agarwal_Zaki_JFM2014,Henningson_Lundbladh_JFM1993} with modifications made in \cref{eq:perturb3D_2,eq:perturb3D_4}.
% \begin{align}
%   \label{eq:perturb3D_1}%
%   \Psi = Kf(y)\left(\frac{x^\prime}{l_x}\right)z^\prime
%   	\exp\left[ -\left(\frac{x^\prime}{l_x}\right)^2 -\left(\frac{z^\prime}{l_z}\right)^2 \right],
% \\
%   \label{eq:perturb3D_2}%
%   (v_x^p,v_y^p,v_z^p) = \left(-\frac{\partial \Psi}{\partial y}\sin\theta,\frac{\partial \Psi}{\partial z^\prime},-\frac{\partial \Psi}{\partial y}\cos\theta\right),
% \\
%   \label{eq:perturb3D_3}%
%   (x^\prime,z^\prime) = (x\cos\theta - z\sin\theta,x\sin\theta + z\cos\theta),
% \\
%   \label{eq:perturb3D_4}%
%   f(y) = (1+y)^2\exp\left(\beta_f(1+y)^2\right)
%   -(1-y)^2\exp\left(\beta_f(1-y)^2\right),
%   % \begin{aligned}
%   % f(y) = &(1+y)^2\exp(\beta_f(1+y)^2)\\
%   % &-(1-y)^2\exp(\beta_f(1-y)^2),
%   % \end{aligned}
% \end{align}
% where $(x,y,z)$ is the spatial coordinates relative to the geometry center. The streamwise and spanwise lengths are controlled by $l_x$ and $l_z$ which are fixed to $2$.  
% A anti-symmetric wall-normal dependence $f(y)$ is chosen. Here,
% $\beta_f$ is adjusted to set the peaks of $f(y)$ at $y\approx \pm 0.7$. $K$ is the magnitude of disturbance.
% and is changed to access different transition pathways. 
For 2D flow, $z$ in \cref{eq:perturb3D_3} is set to $0$. % to acquire a 2D distance.
We emphasize that our reported instability behaviors do not rely on this particular disturbance form.
This was confirmed by testing another form of disturbance.
In particular, at sufficiently small disturbance magnitudes, both forms lead to the same linear instability.

% \section{Results and discussion}\label{Sec_result}

\begin{figure}
	\centering		
	\includegraphics[width=.49\linewidth, trim=0mm 0mm 0mm 0mm, clip]{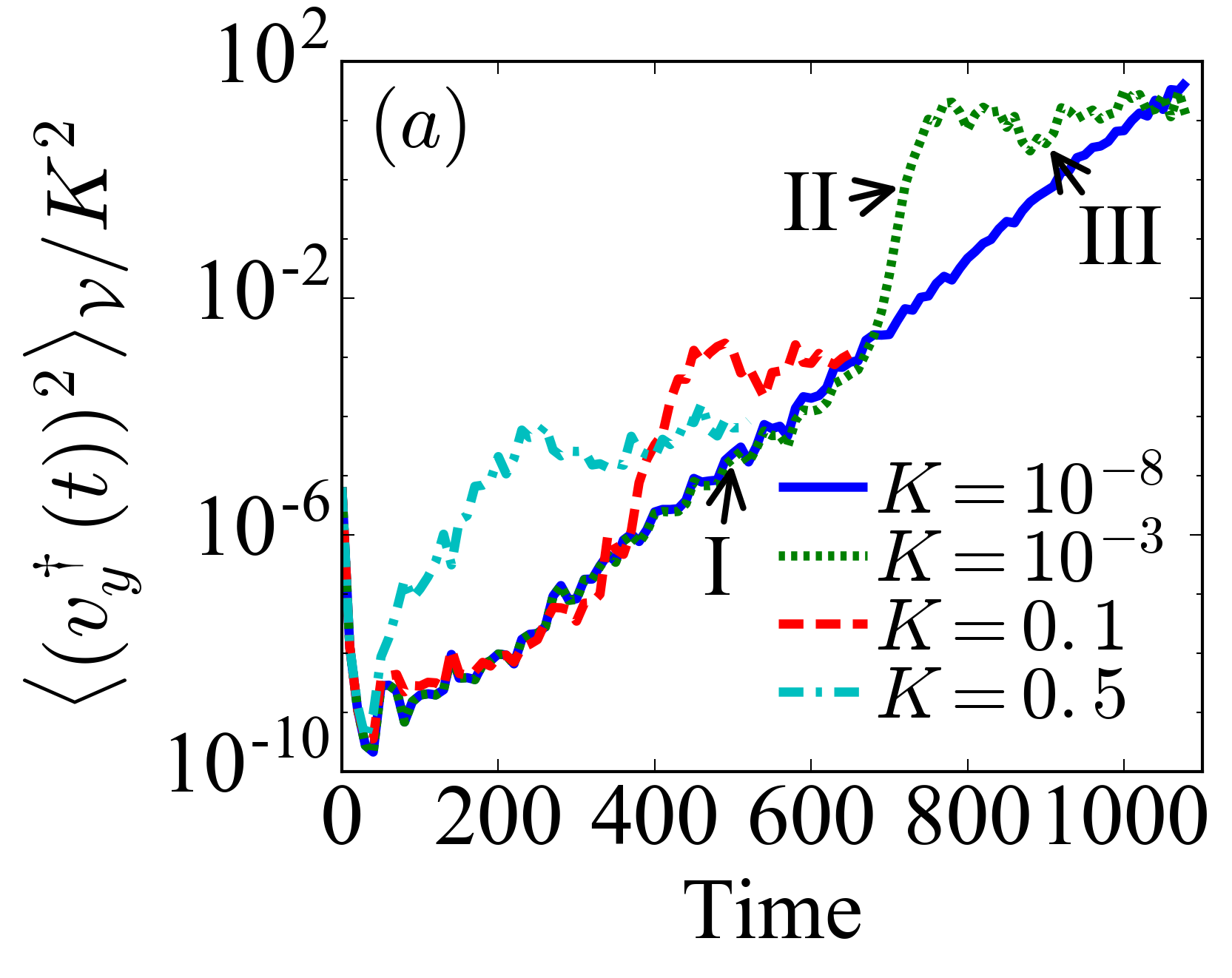}	
	\includegraphics[width=.49\linewidth, trim=0mm 0mm 0mm 0mm, clip]{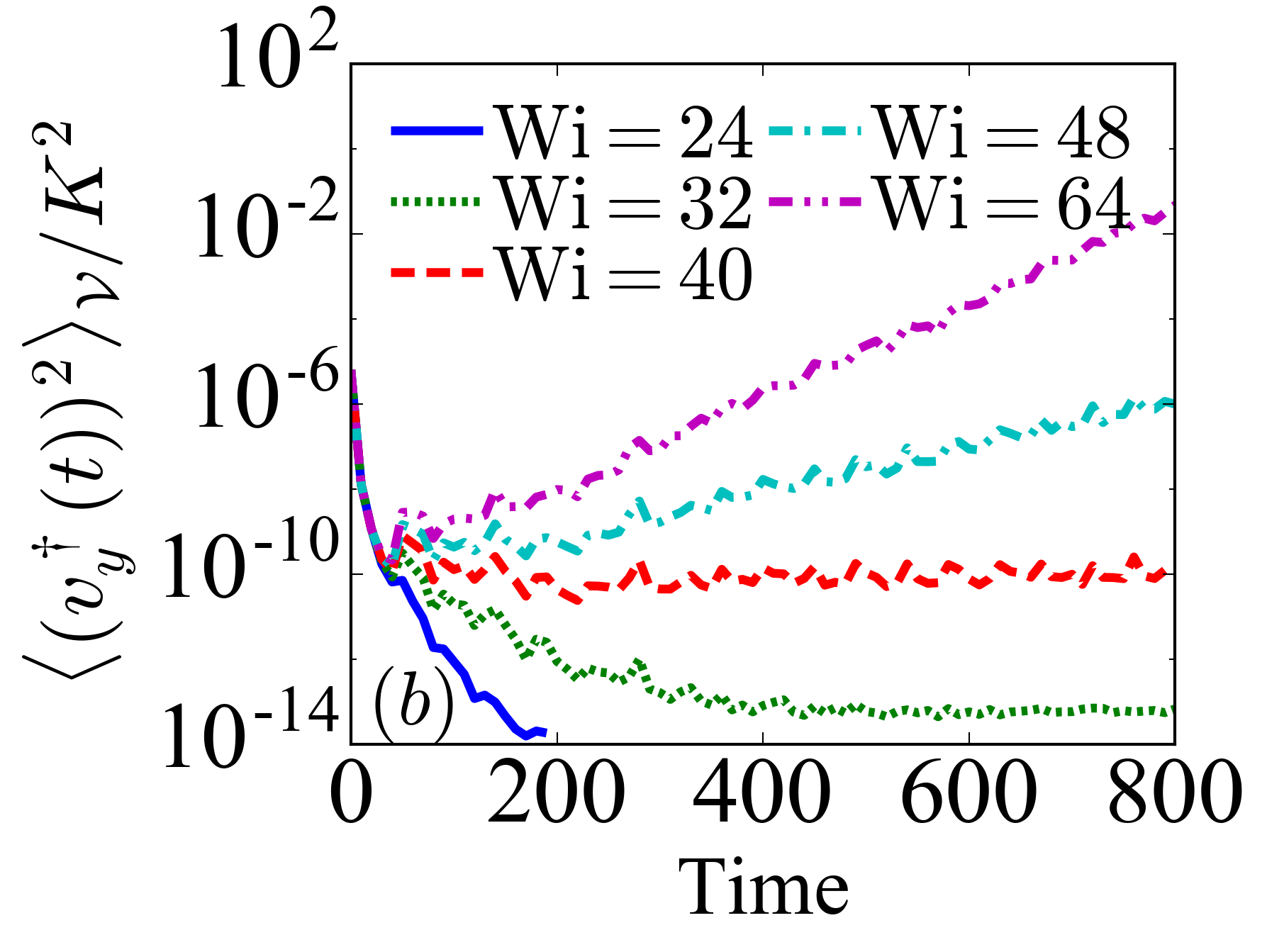}
	\caption{Evolution of velocity disturbance from the perturbed laminar state in 2D at $\mathrm{Re}=3600$ with (a) varying $K$ at $\mathrm{Wi}=64$ and (b) varying $\mathrm{Wi}$ at $K=\num{e-8}$.}
	% Transition pathways in viscoelastic turbulence revealed by time series of $\langle (v_{y}^{\dagger}(t))^2 \rangle_{\mathcal{V}}$: $(a)$ varying $K$ at fixed $(\mathrm{Re},\mathrm{Wi})=(3600,64)$; $(b)$ varying $\mathrm{Wi}$ at fixed $(\mathrm{Re},K)=(3600,10^{-8})$.}
	\label{fig:vy_TS}
\end{figure}

\Cref{fig:vy_TS}(a) shows the growth trajectories of the mean-square wall-normal velocity disturbance $\langle (v_{y}^{\dagger}(t))^2 \rangle_{\mathcal{V}}$ ($\langle\cdot\rangle_{\mathcal{V}}$ denotes volume average) in high-$\mathrm{Wi}$ 2D flow with different initial disturbance magnitudes.
The disturbance grows at all $K$ values tested, down to $K=\num{e-8}$ at which the maximum velocity disturbance is ten orders of magnitude smaller than the base-flow velocity.
Growth of infinitesimal disturbances is the defining feature of linear instability.
At $K=\num{e-8}$, the disturbance grows exponentially with time (\cref{fig:vy_TS}(a) is semilogarithmic), which also confirms the existence of a linear inertioelastic instability (LIEI). We call it inertioelastic because both high $\mathrm{Re}$ and high $\mathrm{Wi}$ are required.
(A small dip is seen at the short-time limit because the arbitrarily-chosen initial disturbance does not match the unstable eigenmode of the instability.)
At $K=\num{e-3}$ and $K=0.1$, the initial growth is still exponential.
With $\langle (v_{y}^{\dagger}(t))^2 \rangle_{\mathcal{V}}$ scaled by $K^2$, growth trajectories over a wide range of $K$ (\numrange{e-8}{e-1}) perfectly collapse in this exponential growth stage -- i.e., $v_y^\dagger(t)\propto K$, which, once again, proves the linearity of the instability.
Departure from exponential growth occurs later when the disturbance reaches the threshold for nonlinear effects.
For the $K=\num{e-3}$ and $K=0.1$ cases shown, the departure occurs at about the same $\langle (v_{y}^{\dagger}(t))^2 \rangle_{\mathcal{V}}$ magnitude of $\sim\mathcal{O}(\num{e-9})$.
For a larger $K=0.5$, the initial disturbance is large enough to trigger direct transition to the nonlinear inertioelastic instability (NIEI) without a noticeable exponential stage.
After the nonlinear growth stage, all cases converge to a plateau corresponding to the EIT state previously found in 2D~\citep{Sid_Terrapon_PRFluids2018,Zhu_Xi_JNNFM2020,Zhu_Xi_PRFluids2021}.

\Cref{fig:vy_TS}(b) shows that, at $\mathrm{Re}=3600$, the linear instability only exists at $\mathrm{Wi}>40$, below which the disturbance decays. Interestingly, this critical $\mathrm{Wi}$ coincides with the minimal $\mathrm{Wi}$ for self-sustaining EIT to emerge in our earlier 3D DNS at the same parameters~\citep{Zhu_Xi_PRFluids2021}.

\begin{figure}
	\centering		
	\includegraphics[width=.99\linewidth, trim=0mm 0mm 0mm 0mm, clip]{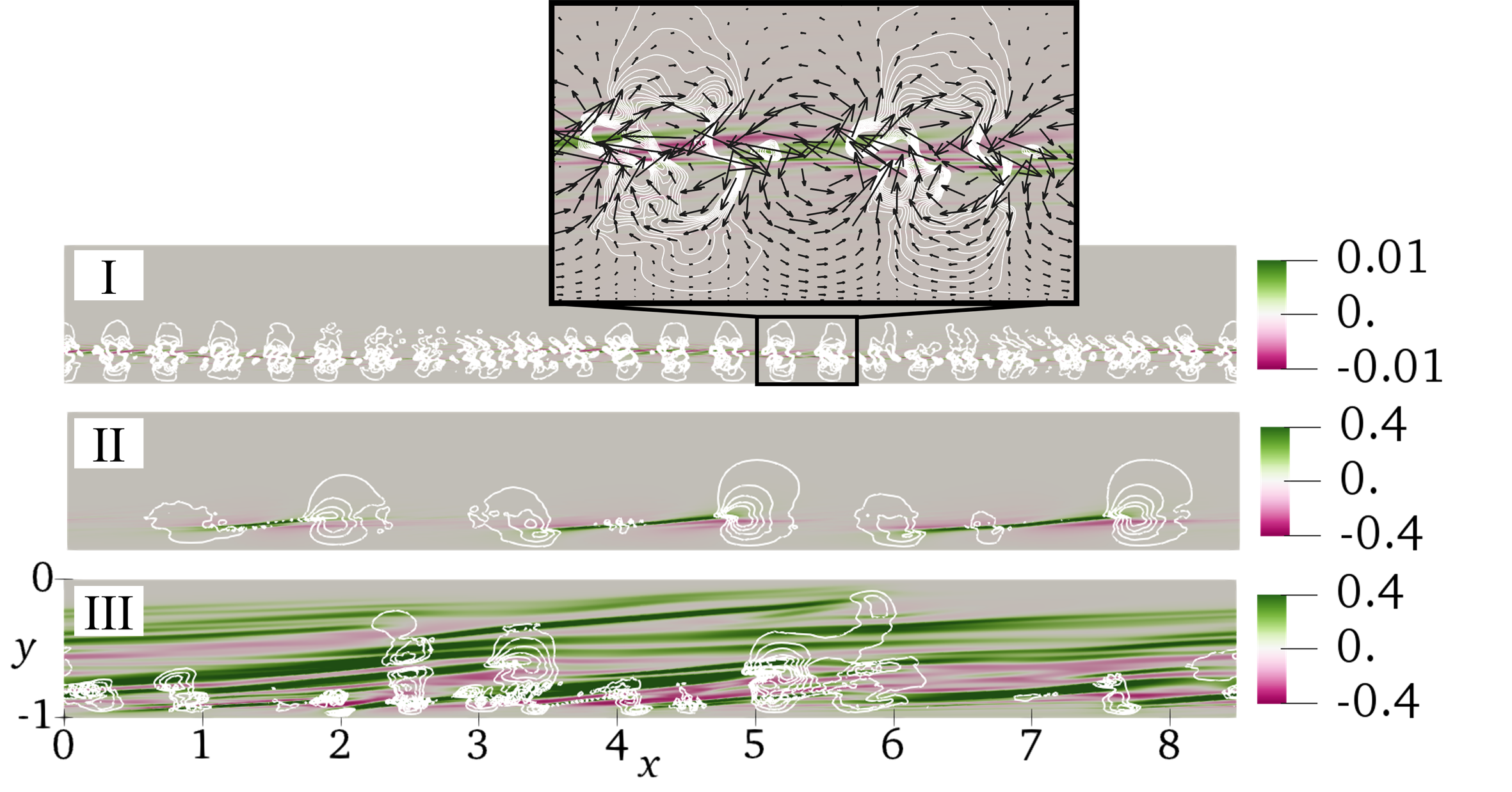}	
	\caption{
	Different stages of instability development (as marked in \cref{fig:vy_TS} for $K=\num{e-3}$) in 2D flow (bottom half channel).
	% Growing 2D instability in the bottom half channel at instants I, II, and III marked in \cref{fig:vy_TS} with $(\mathrm{Re},\mathrm{Wi},K)$=$(3600,64,10^{-3})$.
	Lines represent vortex strength $Q$ contour levels~\citep{Hunt_Wray_CTR1988,Zhu_Xi_JFM2019} (I: $\num{2e-5}\sim\num{4e-4}$; II: $0.002\sim 0.02$, III: $0.005\sim 0.05$).
	Colors map to the $\mathrm{tr}(\alpha)/b$ disturbance field.
	Arrows in the inset show $x$ and $y$ velocity disturbances.}
	% Instantaneous flow fields of 2D DNS (bottom half channel; $(\mathrm{Re},\mathrm{Wi},K)$=$(3600,64,10^{-3})$). Lines represent vortices identified by $Q$-criterion~\citep{Hunt_Wray_CTR1988,Zhu_Xi_JFM2019}. Colors maps to the deviation of $\mathrm{tr}(\alpha)/b$ to the laminar state. }
	\label{fig:flowfield_2d}
\end{figure}

Typical flow structures for instability growth from the LIEI are shown in \Cref{fig:flowfield_2d}.
During the linear (exponential-growth) stage (I), thin inclined polymer sheets are closely attached to the wall, with the maximum $\mathrm{tr}(\alpha)$ disturbance found
% Intensity of the sheets varies and usually achieve a maximum
at $y\approx\pm 0.8$. 
The sheets are sandwiched by two parallel trains of counter-rotating spanwise vortices.
% Induced by the polymer sheets, two parallel trains of vortices are formed on both side of the sheet.
This is clearly a WM instability that differs from the CM instability from linear stability analysis~\citep{Garg_Subramanian_PRL2018}.
The structure is quasi-periodic with some level of chaos.
Over the domain length $L_x$, one can count $\approx 23$ periods, which translates to a wavenumber $k=2\pi/L_x\times 23\approx 17$.
Previous studies mostly explored instabilities at $k\lesssim\mathcal{O}(1)$. The new LIEI occurs at a much smaller length scale than those previously searched.
% The unstable wall mode observed are qualitatively different from the periodic center mode from the linear stability analysis~\citep{Garg_Subramanian_PRL2018} which suggests the existence of multiple unstable mode in viscoelastic turbulence.
% The different unstable modes are indeed observed in the very recent experimental study~\citep{choueiri2021experimental}.
The small linear structures reorganize into $k=\mathcal{O}(1)$ structures with larger vortices and longer polymer sheets during nonlinear growth (II),
% As the flow steps into the busting period (instant II), the polymer sheets extend and stack up, forming large-scale vortices at the leading-edge of the sheets.
Eventually, as $\langle (v_{y}^{\dagger}(t))^2 \rangle_{\mathcal{V}}$ plateaus (III), the solution converges to the steady 2D EIT solution~\citep{Sid_Terrapon_PRFluids2018,Zhu_Xi_JNNFM2020}, where the polymer sheets extend downstream and stack up in a lasagna configuration.
% the polymer sheets will expand the entire channel to form a ``lasagna''-like structures (instant III) where the steady 2D EIT is reached. 
This completes a pathway for EIT growth from a linear instability, which, to our knowledge, was never reported before.

\begin{figure}
	\centering				
	\includegraphics[width=.99\linewidth, trim=0mm 0mm 0mm 0mm, clip]{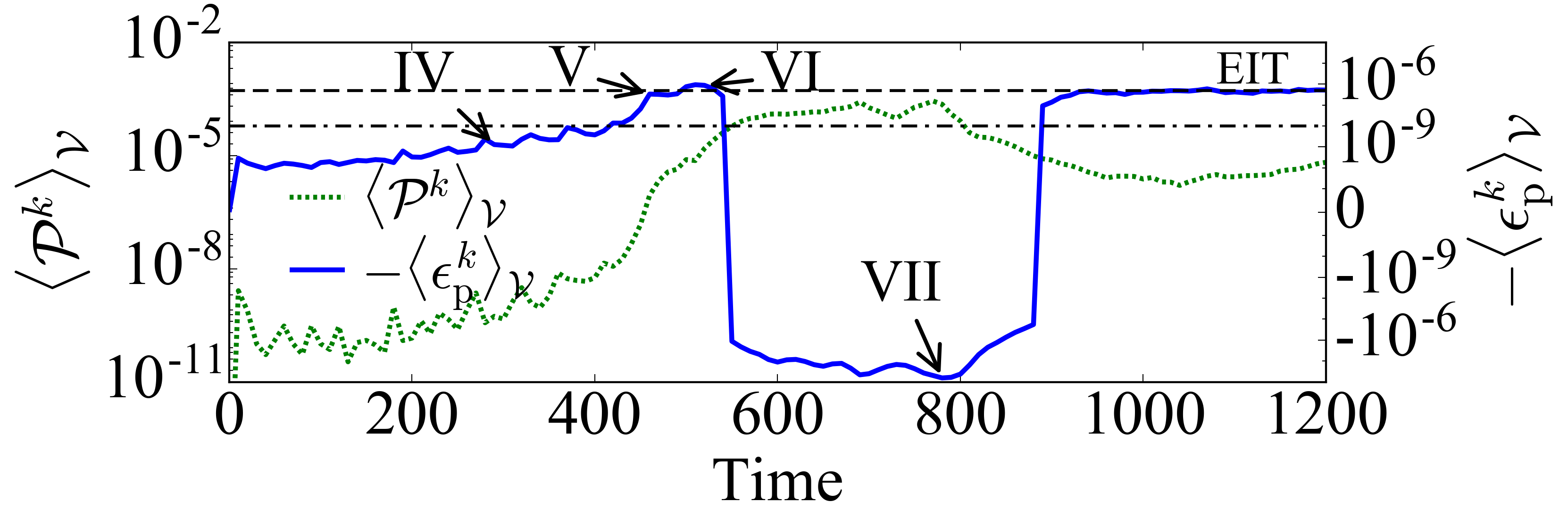}	
	\caption{Contributions to TKE growth from inertial ($\langle\mathcal{P}^k\rangle_\mathcal{V}$) and elastic ($-\langle\epsilon_\text{p}^k\rangle_\mathcal{V}$) mechanisms in 3D flow following the LIEI ($K=0.05$, $\mathrm{Re}=3600$, and $\mathrm{Wi}=64$).
	% 3D instability growth from (a) LIEI ($K=0.05$), (b) NIEI ($K=0.3$), and (c) NIDI ($K=5$), at $\mathrm{Re}=3600$ and $\mathrm{Wi}=64$.
	Dot-dashed and dashed lines mark the $-\langle\epsilon_\text{p}^k\rangle_\mathcal{V}$ magnitudes for the onset of NIEI and steady-state EIT, respectively.}
	% of 3D DNS with different $K$ $(\mathrm{Re}=3600,\mathrm{Wi}=64)$: (a) LEDI ($K=${\color{gray}$0.01$}, {\color{blue}$0.05$}, and {\color{gray}$0.1$}), (b) NEDI ($K=0.3$), (c) NIDI ($K=5$).}	
	\label{fig:tkebudget}
\end{figure}

The same LIEI, showing exponential growth of velocity disturbance in proportion to $K$, is also found in 3D flow down to the lowest $K=\num{e-8}$ tested.
\Cref{fig:tkebudget} shows the inertial and elastic driving forces for turbulence growth from the LIEI using a
% the production of turbulent kinetic energy (TKE) $\left\langle\mathcal{P}^k\right\rangle_\mathcal{V}$ and $-\left\langle\epsilon_\text{p}^k\right\rangle_\mathcal{V}$ after different initial disturbance magnitudes.
turbulent kinetic energy (TKE)  $\langle k\rangle_\mathcal{V}\equiv(1/2)\langle\mbf v'\cdot\mbf v'\rangle_\mathcal{V}$
(``$\prime$'' indicates fluctuations w.r.t. the ensemble average)
balance~\citep{Zhu_Xi_JNNFM2019,Xi_POF2019,Zhu_Xi_PRFluids2021}:
% For 3D DNS, the inertial turbulent structures are able to grow which induce new transition stages and pathways. In \cref{fig:tkebudget}, the balance of average kinetic energy (TKE) $\langle k\rangle_\mathcal{V}\equiv(1/2)\langle\mbf v'\cdot\mbf v'\rangle_\mathcal{V}$
% (``$\prime$'' indicates fluctuating components)~\citep{Zhu_Xi_JNNFM2019,Xi_POF2019,zhu2021nonasymptotic}
\begin{eqnarray}
	\frac{\partial\langle k\rangle_\mathcal{V}}{\partial t}
		= \left\langle\mathcal{P}^k\right\rangle_\mathcal{V}
			-\left\langle\epsilon_\text{v}^k\right\rangle_\mathcal{V}
			-\left\langle\epsilon_\text{p}^k\right\rangle_\mathcal{V}
	\label{eq:tkebal}
\end{eqnarray}
% is used to demonstrate different transition pathways in 3D DNS. Here,
where $\mathcal{P}^k\equiv-\langle v_x'v_y'\rangle(\partial\langle v_x\rangle/\partial y)$
is the TKE production by inertia, 
$\epsilon_\text{v}^k\equiv(2\beta/\mathrm{Re})\langle\mbf{\Gamma}':\mbf{\Gamma}'\rangle$ is its viscous dissipation,
and
$\epsilon_\text{p}^k\equiv(2(1-\beta)/(\mathrm{ReWi}))\langle\mbf{\tau}_\text{p}':\mbf{\Gamma}'\rangle$ is TKE conversion to elastic energy
% are TKE losses by viscous dissipation and conversion to elastic energy, respectively
($\mbf\Gamma\equiv(1/2)(\mbf\nabla\mbf v+(\mbf\nabla\mbf v)^\mathrm{T})$; $\langle\cdot\rangle$ denotes $xz$-average).
Linear growth occurs at $t\lesssim 440$, which is mostly driven by elasticity as $-\langle\epsilon_\text{p}^k\rangle_\mathcal{V}$ is positive and significantly higher than $\langle\mathcal{P}^k\rangle_\mathcal{V}$.
% In fig 3(a), what is the time stamps for (1) onset of nonlinear instability, (2) convergence to EIT, and (3) transition to IDT? I read 430, 460, and 550, respectively from the graph but would like to confirm.
% Yes, they are. I measure the time more carefully, and they are (1) 438, (2) 464, and (3) 542
NIEI occurs at $t\approx 440$, where sudden jumps are found in both $-\langle\epsilon_\text{p}^k\rangle_\mathcal{V}$ and $\langle\mathcal{P}^k\rangle_\mathcal{V}$.
With varying $K$ (not shown), the onset time of the nonlinear departure changes, but it occurs at the same threshold magnitude of $-\langle\epsilon_\text{p}^k\rangle_\mathcal{V}$ (dot-dashed line in \cref{fig:tkebudget}).
At $t\approx 465$, $-\langle\epsilon_\text{p}^k\rangle_\mathcal{V}$ reaches a plateau that matches 3D EIT found previously~\citep{Zhu_Xi_JNNFM2020,Zhu_Xi_PRFluids2021}.
% At this $Wi$, EIT does not persist~\citep{Zhu_Xi_PRFluids2021}.
At $t\approx 540$, $-\langle\epsilon_\text{p}^k\rangle_\mathcal{V}$ quickly drops to negative -- elasticity starts to suppress turbulence, leaving inertia as the only driving force for instability.
This period of IDT is temporary and the flow eventually returns to EIT as its steady state.
At sufficiently high $K$, the LIEI is bypassed to allow direct NIEI or, at larger $K$, direct transition to IDT.
We will call the latter a nonlinear inertia-driven instability (NIDI), where a rapid rise in $\langle\mathcal{P}^k\rangle_\mathcal{V}$ and negative $-\langle\epsilon_\text{p}^k\rangle_\mathcal{V}$ follow the initial disturbance.

\begin{figure}
	\centering				
	\includegraphics[width=.99\linewidth, trim=0mm 0mm 0mm 0mm, clip]{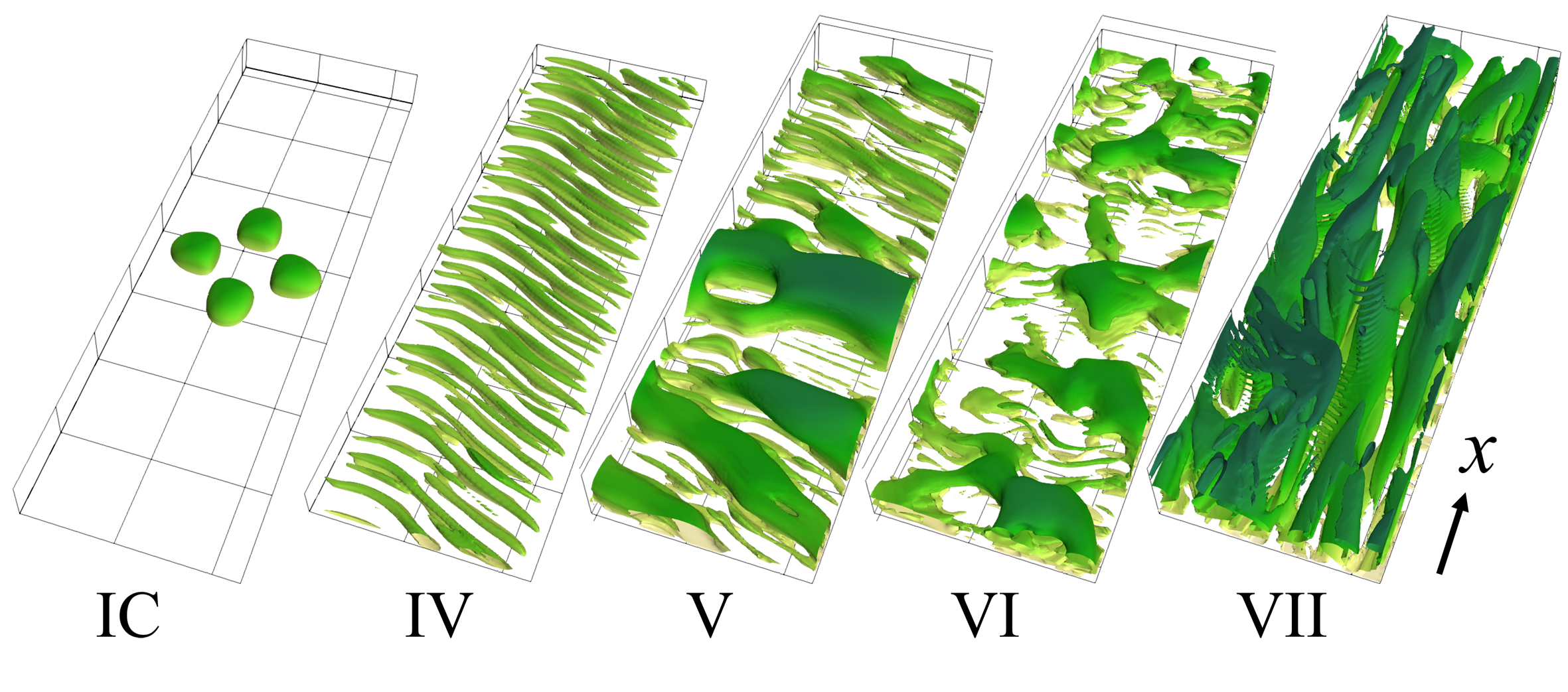}	
	\caption{
	Vortex evolution from the initial condition (IC) to different stages following the LIEI (as marked in \cref{fig:tkebudget}) in 3D flow (bottom half channel).
	Isosurfaces show vortex strength~\citep{Hunt_Wray_CTR1988,Zhu_Xi_JFM2019} ($Q=0.0005$ for IC and $Q=0.005$ for the rest).}
	% Isosurfaces show vortices identified by $Q$-criterion~\citep{Hunt_Wray_CTR1988,Zhu_Xi_JFM2019} ($Q=0.005$ for VI, V, VI, VII, and $Q=0.001$ for VIII), only bottom half channel are presented. Colors maps to distance to the bottom wall. }
	\label{fig:flowfield_3d}
\end{figure}

\begin{figure*}
	\centering				
	\includegraphics[width=.32\linewidth, trim=0mm 0mm 0mm 0mm, clip]{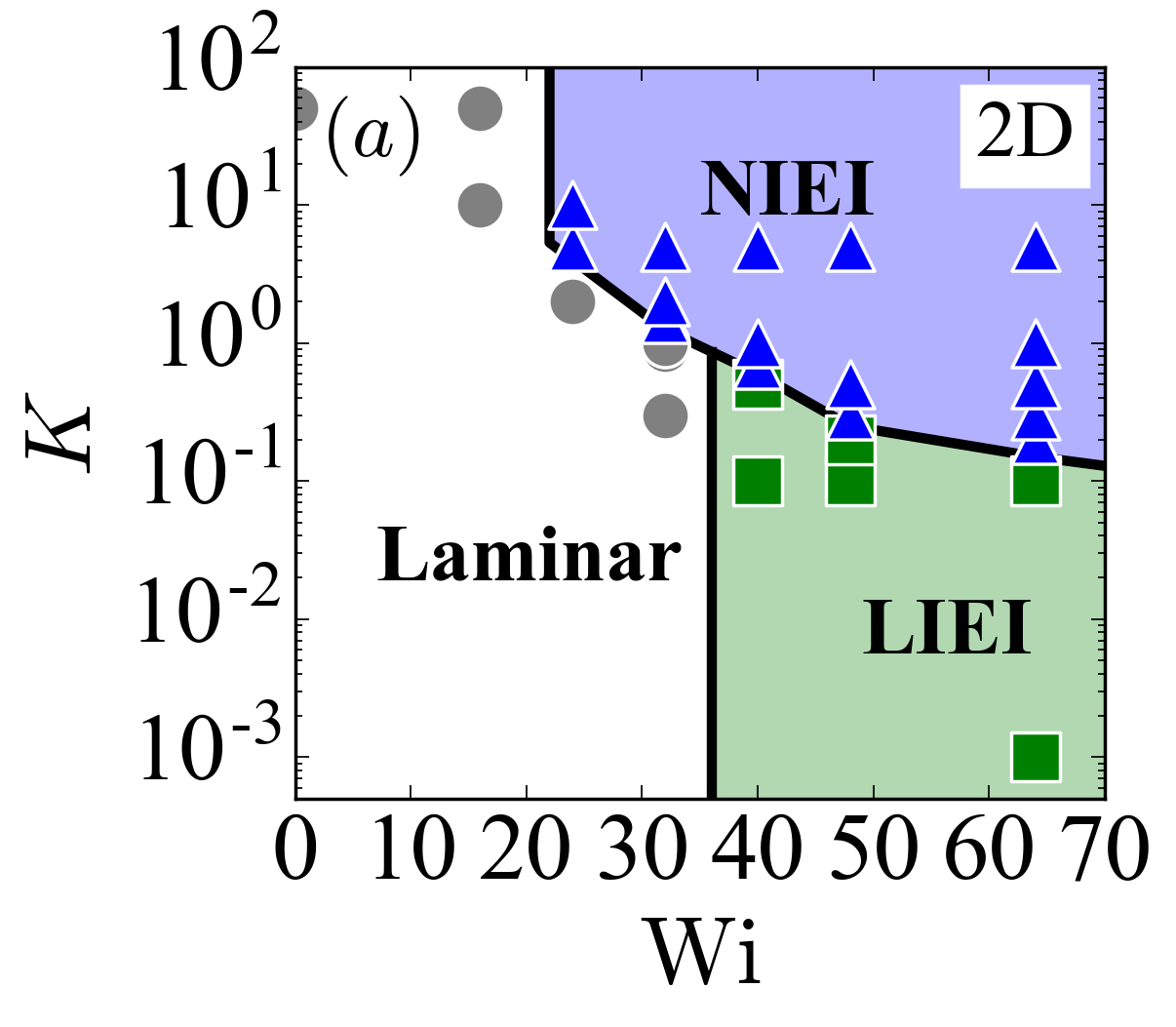}	
	\includegraphics[width=.32\linewidth, trim=0mm 0mm 0mm 0mm, clip]{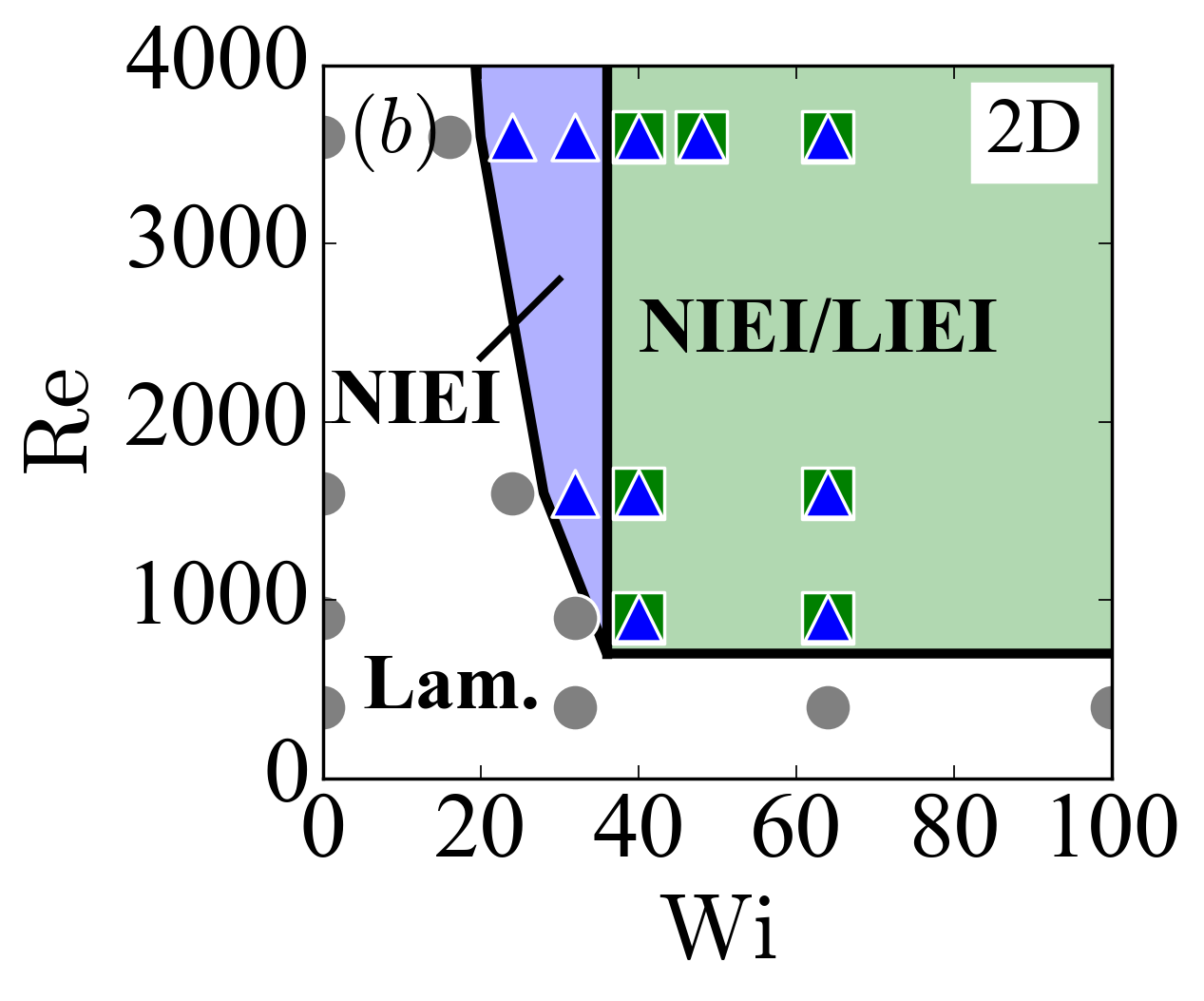}
	\includegraphics[width=.32\linewidth, trim=0mm 0mm 0mm 0mm, clip]{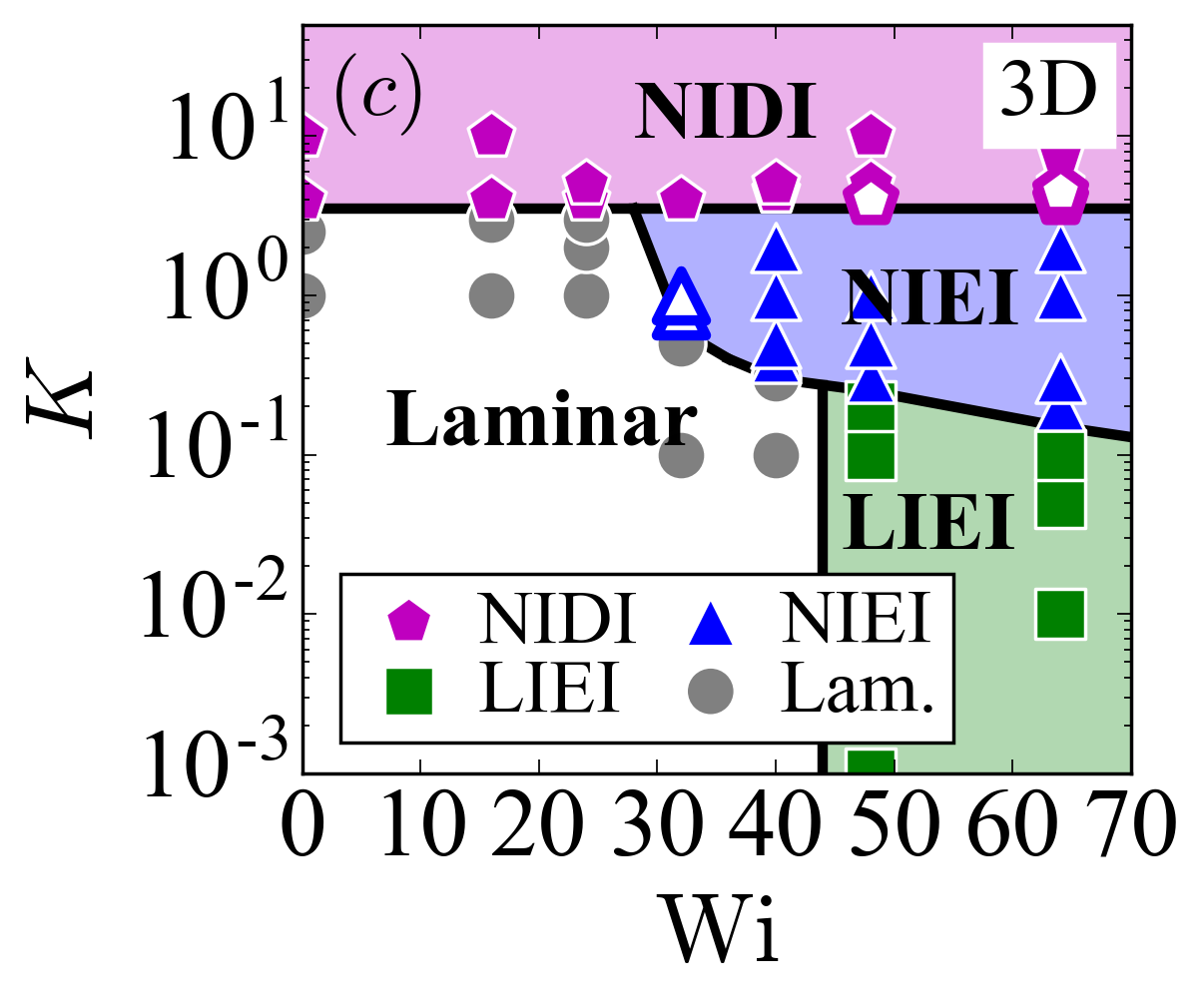}	
	\caption{Instability scenarios: (a) 2D flow in the $K$-$\mathrm{Wi}$ space ($\mathrm{Re}=3600$); (b) 2D flow in the $\mathrm{Re}$-$\mathrm{Wi}$ space ($K=\num{e-8}$ for LIEI; $K=5$ for NIEI); (c) 3D flow in the $K$-$\mathrm{Wi}$ space ($\mathrm{Re}=3600$).
	% Each region shows the first occurring instability.
	Empty symbols indicate that the resulting turbulent state lasts for $<400$ time units.}
	% If the resulting turbulent dynamics is transient, its lower boundary is marked with a dashed line.}
	% transition diagram: $K-\mathrm{Wi}$ ($\mathrm{Re}=3600$) of $(a)$ 2D DNS and $(b)$ 3D DNS and $(c)$ $\mathrm{Wi}-\mathrm{Re}$ ($K=10^{-8}$ and $5$) spaces of 2D DNS.}
	\label{fig:phase_space}
\end{figure*}

\Cref{fig:flowfield_3d} shows the evolving vortex configuration at different stages in \cref{fig:tkebudget}.
During the linear growth stage (IV), the imposed, highly localized disturbance (IC) quickly transforms to narrow, finger-like vortices spreading across the near-wall layer.
The structure closely resembles that of linear growth in 2D (\cref{fig:flowfield_2d}(I)), except for a small tilt in $z$.
Our test with another disturbance form results in a different, but also weak, $z$-dependence. We thus conclude that the $z$-dependence is insignificant and the instability is still fundamentally 2D.
The onset of NIEI is marked by the emergence of large spanwise rolls (V).
The rolls initially reflect their 2D origin but quickly become $z$-dependent as EIT is reached (VI).
Separation from 2D dynamics is expected. Indeed, 3D features of EIT are critical for MDR behaviors~\citep{Zhu_Xi_PRFluids2021}.
During the transient IDT period (VII), spanwise rolls are replaced by streamwise vortices.

We now explore instability behaviors in broader $\mathrm{Wi}$ and $\mathrm{Re}$ ranges.
For 2D flow at $\mathrm{Re=3600}$ (\cref{fig:phase_space}(a)), the LIEI occurs at $\mathrm{Wi}\gtrsim 40$.
The critical $\mathrm{Wi}$ is independent of $K$, as expected for a linear instability.
(Although not included in \cref{fig:phase_space}, all LIEI cases are tested for $K=\num{e-8}$.)
Larger disturbances trigger direct transition to the NIEI, which is found for $\mathrm{Wi}$ down to $24$.
% At lower $\mathrm{Wi}$, the NIEI does not lead to persistent turbulence.
% \textcolor{red}{%
% (Turbulence is considered persistent if it lasts for $>100$ time units (TUs) for EIT or $>1000$ TUs for IDT.)
% }%
\Cref{fig:phase_space}(b) shows that the critical $\mathrm{Wi}$ for the LIEI is also constant over varying $\mathrm{Re}$.
For comparison, the original experiments by \citet{Samanta_Hof_PNAS2013} reported that for the same polymer solution, transition to EIT occurs at the same shear rate (i.e., same $\mathrm{Wi}$) and does not depend on the disturbance magnitude.
The NIEI is found in the same region when a larger disturbance is used, but it also extends to lower $\mathrm{Wi}$, which indicates a subcritical bifurcation.
The minimal $\mathrm{Wi}$ for NIEI decreases with increasing $\mathrm{Re}$.
The lowest $\mathrm{Re}$ where we find instability, linear or nonlinear, is $900$.
The existence of a finite-$\mathrm{Re}$ threshold indicates the necessity of inertia in the instabilities, despite the seemingly more important role of elasticity.
For this reason, the instabilities are considered inertioelastic.

For 3D flow, the $\mathrm{Wi}$-dependence of LIEI and NIEI (\cref{fig:phase_space}(c)) is similar to the 2D case, except that the critical $\mathrm{Wi}$ is slightly higher.
This is likely due to the coarser resolution used in 3D DNS, which
under-resolves the stress shock fronts~\citep{Zhu_Xi_JNNFM2020}.
Larger disturbances cause the NIDI. Its threshold $K$ appears insensitive to increasing $\mathrm{Wi}$, which supports the earlier finding that the laminar-IDT boundary is invariant with $\mathrm{Wi}$~\citep{Xi_Graham_PRL2012,Xi_Bai_PRE2016}.
At high $\mathrm{Wi}$, however, the resulting IDT only exists as short transients (replaced by EIT later).
Shortening of IDT periods with increasing $\mathrm{Wi}$ reflects the turbulence-suppression role of elasticity in that regime, whereas for NIEI, the threshold $K$ decreases with $\mathrm{Wi}$ -- i.e., elasticity is a destabilizing force.
Coexisting inertial and inertioelastic transitions were also reported by \citet{Samanta_Hof_PNAS2013} in their low polymer concentration cases: 
transition to EIT occurs when no extra disturbance was introduced, but with larger disturbances the transition behavior matches that of IDT.

Our most important finding is that EIT can be triggered directly from a WM linear instability of the laminar state.
In this statement, we restrict the term ``EIT'' to the dominant flow type of MDR and other high-$\mathrm{Wi}$ turbulent states in experimentally relevant parameter regimes for the DR problem, while recognizing the possibility that other linear instabilities, such as the CM instability~\citep{Garg_Subramanian_PRL2018,Khalid_Shankar_JFM2021}, could lead to other, so-far unknown, self-sustaining dynamics that are also inertioelastic in nature.
The finding contradicts the prevalent notion that transition to turbulence in viscoelastic fluids, like in Newtonian fluids, must be subcritical.
It also clearly portrays the relationship between different flow states and their relative stability, which paves the way for fully decoding the dynamics of high-$\mathrm{Wi}$ viscoelastic turbulence.

\begin{acknowledgments}
\emph{Acknowledgments} -- The authors acknowledge the financial support from the Natural Sciences and Engineering Research Council of Canada (NSERC; Nos.~RGPIN-2014-04903 and RGPIN-2022-04720) and the allocation of computing resources awarded by the Digital Research Alliance of Canada (www.alliancecan.ca).
The computation is made possible by the facilities of the Shared Hierarchical Academic Research Computing Network (SHARCNET: www.sharcnet.ca).
Our viscoelastic DNS code is based on the Newtonian code \texttt{ChannelFlow 2.0} (https://www.channelflow.ch) developed by John Gibson, Tobias M. Schneider and co-workers.
\end{acknowledgments}

% Note the spaces between the initials
\bibliography{FluidDyn,Polymer,General,Zhu_bibtex}

\end{document}